\documentclass{article}
\usepackage{amsfonts}

\usepackage{graphicx}
\usepackage{amsmath}

\newenvironment{proof}[1][Proof]{\textbf{#1.} }{\ \rule{0.5em}{0.5em}}
\input{tcilatex}

\begin{document}

\title{Boson-Fermion unification \\
implemented by Wick calculus}
\author{John Gough \\
Department of Computing \& Mathematics\\
Nottingham-Trent University, Burton Street,\\
Nottingham NG1\ 4BU, United Kingdom.\\
john.gough@ntu.ac.uk}
\date{}
\maketitle

\begin{abstract}
We construct a transformation between Bose Fock space $\Gamma _{+}\left( 
\frak{h}\right) $ and Fermi Fock space $\Gamma _{-}\left( \frak{h}\right) $
that is super-symmetric in the sense that it converts Boson fields into
Fermi fields over a fixed one-particle space $\frak{h}$. The transformation
the spectral splitting of $\frak{h}$ into a continuous direct integral of
internal spaces $\left( \frak{k}_{\omega }\right) _{\omega }$. We present a
theory of integration on the Fock spaces over $L^{2}\left( \left( \frak{k}%
_{\omega }\right) _{\omega },\mathbb{R}_{+},d\omega \right) $ that is a
natural generalization of the theory of quantum stochastic calculus and
which we refer to as a Wick calculus.
\end{abstract}


\section{Introduction}

The formalism of second quantization is fundamental to modern physics \cite
{Berezin} and admits a natural functional calculus. A very specialized
version of this is quantum stochastic calculus. Quantum stochastic
processes, parameterized by time $t,$ are families of operators on Fock
space over Hilbert spaces of the type $L^{2}\left( \frak{k},\mathbb{R}%
_{+},dt\right) $. That is, $\frak{k}$ is a fixed Hilbert space, termed the 
\emph{internal space}, and $\phi \in L^{2}\left( \frak{k},\mathbb{R}%
_{+},dt\right) $ is square-integrable $\frak{k}$ valued function. As is
well-known, we have the natural isomorphism $L^{2}\left( \frak{k},\mathbb{R}%
_{+},dt\right) \cong \frak{k}\otimes L^{2}\left( \mathbb{R}_{+},dt\right) $.
The stochastic calculus has been developed in the Boson case \cite{HP:QIFSE}%
, generalizing the classical It\^{o} theory,\ and Fermion setting \cite{AH},
generalizing the Clifford-It\^{o} theory \cite{BSW}. The Bose and Fermi
theories have been unified by means of a continuous version of the
Jordan-Wigner transformation \cite{HP:UFBSC}.

Here we wish to consider second quantizations of Hilbert spaces of the type $%
L^{2}\left( \left( \frak{k}_{\omega }\right) _{\omega },\mathbb{R}%
_{+},d\omega \right) $\ where now we work with families $\left( \frak{k}%
_{\omega }\right) _{\omega }$ of internal spaces indexed by
parameter(interpreted here as frequency) $\omega >0$. The study of Wiener-It%
\^{o} integrals (in the time domain) on Fock spaces over direct integral
Hilbert spaces was first considered by Sunder \cite{Sunder}.

Our motivation comes from modelling physical quantum reservoirs. In such
cases, the $\frak{k}_{\omega }$\ arise as the mass shell Hilbert spaces for
a fixed energy $\omega $.

An infinitely extended quantum reservoir can be considered as the second
quantization of particle having one-particle space $\frak{h}$ and having
one-particle Hamiltonian $H\geq 0$ which is a fixed self-adjoint operator $H$
on $\frak{h}$. Specifically, we take $H$ to have absolutely continuous
spectrum. There then exists an orthogonal projection valued measure $\Pi %
\left[ .\right] $ concentrated on $[0,\infty )$ such that 
\begin{equation}
H\equiv \int_{\lbrack 0,\infty )}\omega \,\Pi \left[ d\omega \right] . 
\tag{1.1}
\end{equation}
Given $\Omega >0$ we consider the subspace $\mathcal{D}_{\Omega }\subset 
\frak{h}$ such that 
\begin{equation}
\int_{-\infty }^{+\infty }dt\ \left| \left\langle f|\exp \left\{ i\left(
H-\Omega \right) t\right\} g\right\rangle \right| <\infty  \tag{1.2}
\end{equation}
whenever $f,g\in \mathcal{D}_{\Omega }$. On this domain we define the
sesquilinear form 
\begin{equation}
\left( g|f\right) _{\Omega }:=\int_{-\infty }^{+\infty }\left\langle f|\exp
\left\{ i\left( H-\Omega \right) t\right\} g\right\rangle \,dt  \tag{1.3}
\end{equation}
and it is convenient to consider the Hilbert space $\frak{k}_{\Omega }$
obtained by factoring out from $\mathcal{D}_{\Omega }$ the null elements $%
\mathcal{N}_{\Omega }=\left\{ f:\left( f|f\right) _{\Omega }=0\right\} $ and
taking the Hilbert space completion with respect to the $\left( .|.\right)
_{\Omega }$-norm. Formally we have 
\begin{equation}
2\pi \,\left\langle f|\,\Pi \left[ d\omega \right] \,g\right\rangle \equiv
\left( g|f\right) _{\omega }\,d\omega  \tag{1.4}
\end{equation}

We then obtain the continuous direct integral decomposition 
\begin{equation}
\frak{h}\cong \int_{\lbrack 0,\infty )}^{\oplus }d\omega \;\frak{k}_{\omega
}.  \tag{1.5}
\end{equation}
For each $\omega \geq 0$, $\frak{k}_{\omega }$ is a Hilbert space with inner
product $\left( .|.\right) _{\omega }$ and we may consider $\frak{h}$ to
consists of all vectors $\phi =\left( \phi _{\omega }\right) _{\omega \geq
0} $, where $\phi _{\omega }\in \frak{k}_{\omega }$ and $\int_{[0,\infty
)}\left( \phi _{\omega }|\phi _{\omega }\right) _{\omega }$ $<+\infty $. The
inner product on $\frak{h}$ may be represented by 
\begin{equation}
\left\langle \phi |\psi \right\rangle =\int_{[0,\infty )}d\omega \;\left(
\phi _{\omega }|\psi _{\omega }\right) _{\omega }.  \tag{1.6}
\end{equation}
We may write $\frak{h}$ as $L^{2}\left( \left( \frak{k}_{\omega }\right)
_{\omega },\mathbb{R}_{+},d\omega \right) $. It is natural, in the light of
the development of quantum stochastic calculus, to consider the space $L_{%
\mathrm{loc}}^{2}\left( \left( \frak{k}_{\omega }\right) _{\omega },\mathbb{R%
}_{+},d\omega \right) $ of locally square-integrable objects $\phi =\left(
\phi _{\omega }\right) _{\omega }$\ where now we only require $%
\int_{B}d\omega \;\left( \phi _{\omega }|\phi _{\omega }\right) _{\omega
}<+\infty $ for any compact Borel subset $B$.

\noindent \textbf{Example: }The basic situation we have in mind \cite{AFL}
is a reservoir particle moving in $\nu $-dimensions and having the spectrum
of elementary excitations $\omega =\omega \left( \mathbf{k}\right) $ where $%
\mathbf{k}=\left( k_{1},\dots ,k_{\nu }\right) $ are the momenta coordinate.
We take $\frak{h}=L^{2}\left( \mathbb{R}^{\nu },d^{\nu }k\right) $ and $%
\left( Hf\right) \left( \mathbf{k}\right) =\omega \left( \mathbf{k}\right)
f\left( \mathbf{k}\right) $. We shall take it that the spectrum foliates the
momentum space into the mass shells $M_{\omega }:=\left\{ \mathbf{k}:\omega
\left( \mathbf{k}\right) =\omega \right\} $ and that the Lesbegue measure on 
$\mathbb{R}^{\nu }$ can be decomposed locally as $d^{\nu }k=d\omega \times
d\sigma _{\omega }$ where $\omega =\omega \left( \mathbf{k}\right) $ and $%
d\sigma _{\omega }$ is surface measure on $M_{\omega }$. (For instance, if $H
$ corresponds to $-\Delta $ in the position representation then $\sqrt{%
\omega \left( \mathbf{k}\right) }$ is radial coordinate and $d\sigma
_{\omega }$ will be surface measure on the sphere of radius $\sqrt{\omega }$%
.) The individual Hilbert spaces are $\frak{k}_{\omega }=L^{2}\left(
M_{\omega },d\sigma _{\omega }\right) $.

Note that we can deal with quasi-free gauge-invariant states on $\Gamma
_{\pm }\left( \frak{h}\right) $ provided only that the covariance matrix $Q$
commutes with the filtration of $\frak{h}$\ induced by $H$. Since we have $%
Q=\coth \frac{\beta H-\mu }{2}$ (Bose) and $Q=\tanh \frac{\beta H-\mu }{2}$
(Fermi) for the free particle Gibbs states, this construction arises
naturally.

\section{Mathematical Notations and Preliminaries}

Let $\Gamma \left( \frak{h}\right) :=\bigoplus_{n=0}^{\infty }\frak{h}%
^{\otimes n}$ be the ``Full'' Fock space over a fixed complex separable
Hilbert space $\frak{h}$. The (anti)-symmetrization operators $\Pi _{\pm }$
are define through linear extension of the relations $\Pi _{\pm
}f_{1}\otimes \cdots \otimes f_{n}$ $:=$ $\frac{1}{n!}\sum_{\sigma \in \frak{%
S}_{n}}\left( \pm \right) ^{\sigma }$ $f_{\sigma \left( 1\right) }\otimes
\cdots \otimes f_{\sigma \left( n\right) }$, with $f_{j}\in \frak{h}$, $%
\frak{S}_{n}$ denotes the permutation group on $\left\{ 1,\dots ,n\right\} $
and $\left( -1\right) ^{\sigma }$ is the parity of the permutation $\sigma $%
. The Bose Fock space $\Gamma _{+}\left( \frak{h}\right) $ and the Fermi
Fock space $\Gamma _{-}\left( \frak{h}\right) $\ having $\frak{h}$ as
one-particle space are then defined as the subspaces $\Gamma _{\pm }\left( 
\frak{h}\right) :=\Pi _{\pm }\,\Gamma \left( \frak{h}\right) $. As usual, we
distinguish the Fock vacuum $\Phi =\left( 1,0,0\dots \right) $, though we
shall write $\Phi _{\pm }$ for emphasis.

Let $g\in \frak{h}$, $U$ unitary and $H$ self-adjoint on $\frak{h}$. We
define the following operators on the Full Fock space 
\begin{eqnarray}
A^{+}\left( h\right) \,f_{1}\otimes \cdots \otimes f_{n} &:&=\sqrt{n+1}%
\;h\otimes f_{1}\otimes \cdots \otimes f_{n};  \notag \\
A^{-}\left( h\right) \,f_{1}\otimes \cdots \otimes f_{n} &:&=\frac{1}{\sqrt{n%
}}\;\left\langle h|f_{1}\right\rangle \,f_{2}\otimes \cdots \otimes f_{n}; 
\notag \\
\Gamma \left( U\right) \,f_{1}\otimes \cdots \otimes f_{n} &:&=\left(
Uf_{1}\right) \otimes \cdots \otimes \left( Uf_{n}\right) ;  \notag \\
\gamma \left( H\right) \,f_{1}\otimes \cdots \otimes f_{n}
&:&=\sum_{j}f_{1}\otimes \cdots \otimes \left( Hf_{j}\right) \otimes \cdots
\otimes f_{n}.  \TCItag{2.1}
\end{eqnarray}
Bose creation and annihilation fields are then defined on $\Gamma _{+}\left( 
\frak{h}\right) $ as 
\begin{equation}
B^{\pm }\left( h\right) :=\Pi _{+}\,A^{\pm }\left( h\right) \,\Pi _{+} 
\tag{2.2}
\end{equation}
while Fermi creation and annihilation fields are defined on $\Gamma
_{+}\left( \frak{h}\right) $ as 
\begin{equation}
F^{\pm }\left( h\right) :=\Pi _{-}\,A^{\pm }\left( h\right) \,\Pi _{-}. 
\tag{2.3}
\end{equation}
Using the traditional conventions $\left[ A,B\right] =AB-BA$ and $\left\{
A,B\right\} =AB+BA$, we have the canonical (anti)-commutation relations 
\begin{equation}
\left[ B^{-}\left( f\right) ,B^{+}\left( g\right) \right] =\left\langle
f|g\right\rangle ;\quad \left\{ F^{-}\left( f\right) ,F^{+}\left( g\right)
\right\} =\left\langle f|g\right\rangle .  \tag{2.4}
\end{equation}
Second quantization operators are defined as $\Gamma _{\pm }\left( U\right)
:=\Pi _{\pm }\,\Gamma \left( U\right) \,\Pi _{\pm }$ and differential second
quantization operators as $\gamma _{\pm }\left( U\right) :=\Pi _{\pm
}\,\gamma \left( U\right) \,\Pi _{\pm }$. We have the relation 
\begin{equation}
\exp \left\{ it\,\gamma _{\pm }\left( H\right) \right\} =\Gamma _{\pm
}\left( e^{itH}\right) .  \tag{2.5}
\end{equation}
More generally we may take the argument of the differential second
quantizations to be bounded: for the rank-one operator $H=\left|
f\right\rangle \left\langle g\right| $ described in standard Dirac bra-ket
notation, we have $\gamma _{+}\left( \left| f\right\rangle \left\langle
g\right| \right) \equiv B^{+}\left( f\right) B^{-}\left( g\right) $ and $%
\gamma _{-}\left( \left| f\right\rangle \left\langle g\right| \right) \equiv
F^{+}\left( f\right) F^{-}\left( g\right) $.The following relations will be
useful 
\begin{equation}
\Gamma _{+}\left( U\right) \,B^{\pm }\left( \phi \right) \,\Gamma _{+}\left(
U^{\dagger }\right) =B^{\pm }\left( U\phi \right) ;\quad \Gamma _{-}\left(
U\right) \,F^{\pm }\left( \phi \right) \,\Gamma _{-}\left( U^{\dagger
}\right) =F^{\pm }\left( U\phi \right)  \tag{2.6}
\end{equation}

In the Bose case, the exponential vector map $\varepsilon :\frak{h}\mapsto
\Gamma _{+}\left( \frak{h}\right) $ is introduced as 
\begin{equation}
\varepsilon \left( f\right) =\oplus _{n=0}^{\infty }\frac{1}{\sqrt{n!}}%
f^{\otimes n}  \tag{2.7}
\end{equation}
with $f^{\otimes n}$ the $n$-fold tensor product of $f$ with itself. The
Fock vacuum is, in particular, given by $\Phi =\varepsilon \left( 0\right) $%
. The set $\varepsilon \left( \frak{h}\right) $ is total in $\Gamma
_{+}\left( \frak{h}\right) $ and we note that $\left\langle \varepsilon
\left( f\right) |\varepsilon \left( g\right) \right\rangle =\exp
\left\langle f|g\right\rangle $. The next result is the basis for the
development of a calculus of second quantized fields, see \cite{Partha} for
proofs.

\bigskip

\noindent \textbf{Lemma (2.1): }\emph{The operations of Bose or Fermi second
quantization have the natural functorial property} 
\begin{equation}
\Gamma _{\pm }\left( \frak{h}_{1}\oplus \frak{h}_{2}\right) \cong \Gamma
_{\pm }\left( \frak{h}_{1}\right) \otimes \Gamma _{\pm }\left( \frak{h}%
_{2}\right) \ .  \tag{2.8}
\end{equation}

\section{Spectral Processes and Wick Calculus}

Let $\frak{h}$\ be a separable complex Hilbert space admitting the
continuous direct integral decomposition 
\begin{equation}
\frak{h}=\int_{[0,\infty )}^{\oplus }d\omega \;\frak{k}_{\omega }.  \tag{3.1}
\end{equation}
That is, for each $\omega \geq 0$, $\frak{k}_{\omega }$ is a Hilbert space
with inner product $\left( .|.\right) _{\omega }$ and we have that $\frak{h}$
consists of all vectors $\phi =\left( \phi _{\omega }\right) _{\omega \geq
0} $, where $\phi _{\omega }\in \frak{k}_{\omega }$ and we have the inner
product on $\frak{h}$%
\begin{equation}
\left\langle \phi |\psi \right\rangle =\int_{[0,\infty )}d\omega \;\left(
\phi _{\omega }|\psi _{\omega }\right) _{\omega }.  \tag{3.2}
\end{equation}
For $0\leq \Omega _{1}<\Omega _{2}$, let $\frak{k}_{[\Omega _{1},\Omega
_{2}]}=\int_{[\Omega _{1},\Omega _{2}]}^{\oplus }d\omega \;\frak{k}_{\omega
} $, then in these notations 
\begin{equation}
\frak{h}\cong \frak{k}_{[0,\Omega ]}\oplus \frak{k}_{(\Omega ,\infty )} 
\tag{3.3}
\end{equation}
for each $\Omega >0$. This leads to the following continuous tensor product
decomposition for Fock space 
\begin{equation}
\Gamma _{+}\left( \frak{h}\right) \cong \Gamma _{+}\left( \frak{k}%
_{[0,\Omega ]}\right) \otimes \Gamma _{+}\left( \frak{k}_{(\Omega ,\infty
)}\right) .  \tag{3.4}
\end{equation}

We define an absolutely continuous, orthogonal projection valued measure $%
\Pi $ on $[0,\infty )$ by 
\begin{equation}
\left( \Pi _{A}\phi \right) _{\omega }:=1_{A}\left( \omega \right) \phi
_{\omega }  \tag{3.5}
\end{equation}
where $A$ is any Borel set and $1_{A}$ its characteristic function.

\bigskip

Next of all, fix an \emph{initial Hilbert space} $\frak{H}_{0}$ and set 
\begin{equation}
\frak{H}:=\frak{H}_{0}\otimes \Gamma _{+}\left( \frak{h}\right) ;\quad \frak{%
H}_{\Omega ]}:=\frak{H}_{0}\otimes \Gamma _{+}\left( \frak{k}_{[0,\Omega
]}\right) ;\quad \frak{H}_{(\Omega }:=\Gamma _{+}\left( \frak{k}_{(\Omega
,\infty )}\right) .  \tag{3.6}
\end{equation}

\bigskip

\noindent \textbf{Definition (3.1):} \emph{A family }$\left( X_{\Omega
}\right) _{\Omega \geq 0}$ \emph{of operators on }$\frak{H}$\emph{\ is said
to be spectrally-adapted if, for each }$\Omega >0$\emph{, the operator }$%
X_{\Omega }$\emph{\ is the algebraic ampliation to }$\frak{H}_{0}\underline{%
\otimes }\varepsilon \left( \frak{k}_{[0,\Omega )}\right) \underline{\otimes 
}\varepsilon \left( \frak{k}_{(\Omega ,\infty )}\right) $\emph{\ of an
operator on }$\frak{H}_{\Omega ]}$\emph{\ with domain }$\frak{H}_{0}%
\underline{\otimes }\varepsilon \left( \frak{k}_{[0,\Omega )}\right) $\emph{%
. We also demand the existence of an adjoint process }$\left( X_{\Omega
}^{\dagger }\right) _{\Omega \geq 0}$\emph{\ having the same ampliation
structure. (Here }$\underline{\otimes }$\emph{\ denotes the algebraic tensor
product.)}

\bigskip

\noindent \textbf{Definition (3.2):} \emph{Let }$\phi \in \frak{h}$\emph{,
we define the (Bosonic) creation and annihilation spectral processes on }$%
\Gamma _{+}\left( \frak{h}\right) $\emph{\ to be} 
\begin{equation}
B_{\phi }^{\pm }\left( \Omega \right) :=B^{\pm }\left( \Pi _{\left[ 0,\Omega %
\right] }\phi \right)  \tag{3.7}
\end{equation}
\emph{and the conservation spectral process to be} 
\begin{equation}
\Lambda \left( \Omega \right) :=\gamma _{+}\left( \Pi _{\left[ 0,\Omega %
\right] }\right)  \tag{3.8}
\end{equation}
\emph{for each }$\Omega \geq 0$\emph{.}

\bigskip

The operators $B_{\phi }^{\pm }\left( \Omega \right) $, $\Lambda \left(
\Omega \right) $ are spectrally-adapted in this sense.

Let $\left( X_{jk}\left( \Omega \right) \right) _{\Omega \geq 0}$ be
piecewise constant, spectrally-adapted processes for $j,k\in \left\{
0,1\right\} $. The \emph{Wick integral} 
\begin{equation}
X_{\Omega }=\int_{[0,\Omega ]}\left( X_{11}\otimes \Lambda \left( d\omega
\right) +X_{10}\otimes B_{\phi }^{+}\left( d\omega \right) +X_{01}\otimes
B_{\psi }^{-}\left( d\omega \right) +X_{00}\otimes d\omega \right)  \tag{3.9}
\end{equation}
is defined in such a way that $\left\langle u\otimes \varepsilon \left(
f\right) |\,X\,v\otimes \varepsilon \left( g\right) \right\rangle $ is
interpreted as 
\begin{equation}
\int_{0}^{\Omega }\left\langle u\otimes \varepsilon \left( f\right)
|\,\left( X_{11}\left( f_{\omega }|g_{\omega }\right) _{\omega
}+X_{10}\left( f_{\omega }|\phi _{\omega }\right) _{\omega }+X_{01}\left(
\psi _{\omega }|g_{\omega }\right) _{\omega }+X_{00}\right) \,v\otimes
\varepsilon \left( g\right) \right\rangle \,d\omega  \tag{3.10}
\end{equation}
for all $u,v\in \frak{H}_{0}$ and $f,g\in \frak{h}$. Formally we write this
as $X_{\Omega }\equiv \int_{\left[ 0,\Omega \right] }X\left( d\omega \right) 
$. Similarly, we set $X_{\Omega }^{\dagger }=$ $\int_{[0,\Omega
]}(X_{11}\otimes \Lambda \left( d\omega \right) $ $+X_{10}\otimes B_{\phi
}^{-}\left( d\omega \right) $ $+X_{01}\otimes B_{\psi }^{+}\left( d\omega
\right) $ $+X_{00}\otimes d\omega )$.

\bigskip

\noindent \textbf{Lemma (3.3): }\emph{Let }$X_{\Omega }$\emph{\ be the
stochastic integral with piecewise constant, spectrally adapted coefficients
as above, then}

\begin{eqnarray}
&&\left\| X_{\Omega }\,u\otimes \varepsilon \left( f\right) \right\|
^{2}\leq \int_{\lbrack 0,\Omega ]}d\omega \;\exp \left\{ \Omega -\omega
+3\int_{\omega }^{\Omega }\left( f_{\omega ^{\prime }}|f_{\omega ^{\prime
}}\right) d\omega ^{\prime }\right\}   \notag \\
&&\times \left[ 3\left( f_{\omega }|f_{\omega }\right) _{\omega }\left\|
X_{11}\left( \omega \right) \,u\otimes \varepsilon \left( f\right) \right\|
^{2}+3\left( \phi _{\omega }|\phi _{\omega }\right) _{\omega }\left\|
X_{10}\left( \omega \right) \,u\otimes \varepsilon \left( f\right) \right\|
^{2}\right.   \notag \\
&&\left. +\left( \psi _{\omega }|\psi _{\omega }\right) _{\omega }\left\|
X_{01}\left( \omega \right) \,u\otimes \varepsilon \left( f\right) \right\|
^{2}+\left\| X_{00}\left( \omega \right) \,u\otimes \varepsilon \left(
f\right) \right\| ^{2}\right] .  \TCItag{3.11}
\end{eqnarray}

\begin{proof}
This is a generic type of estimate in quantum stochastic calculus and in our
case is a straightforward adaptation of section 2 of \cite{HP:SDUCCPS} and
we omit it.
\end{proof}

\bigskip

Let $\left( X_{jk}\left( \Omega \right) \right) _{\Omega \geq 0}$\ be
spectrally adapted processes that are weakly measurable and satisfy the
following locally square-integrability conditions (for arbitrary\emph{\ }$%
u\in \frak{H}_{0}$, $f\in \frak{h}$\emph{)} 
\begin{eqnarray*}
\int_{\lbrack 0,\Omega ]}d\omega \;\left( f_{\omega }|f_{\omega }\right)
_{\omega }\left\| X_{11}\left( \omega \right) \,u\otimes \varepsilon \left(
f\right) \right\| ^{2} &<&\infty ; \\
\int_{\lbrack 0,\Omega ]}d\omega \;\left\| X_{jk}\left( \omega \right)
\,u\otimes \varepsilon \left( f\right) \right\| ^{2} &<&\infty ,\qquad \text{%
otherwise}.
\end{eqnarray*}
Then the Wick integral 
\begin{equation*}
X_{\Omega }=\int_{[0,\Omega ]}\left( X_{11}\otimes \Lambda \left( d\omega
\right) +X_{10}\otimes B_{\phi }^{+}\left( d\omega \right) +X_{01}\otimes
B_{\psi }^{-}\left( d\omega \right) +X_{00}\otimes d\omega \right)
\end{equation*}
exists and is well-defined. It can be understood as the limit of an
approximating sequence $\left( X_{\Omega }^{\left( n\right) }\right)
_{\Omega \geq 0}$, each one constructed using piecewise continuous
coefficients: the approximating coefficients $X_{j}^{\left( n\right) }$
should be chosen so that $\int_{[0,\Omega ]}d\omega \;\left\| \left(
X_{jk}-X_{jk}^{\left( n\right) }\right) \right\| ^{2}\rightarrow 0$ and the
limit process will be independent of the approximating sequence used.

It is useful to use the differential notation $X_{\Omega }\equiv \int_{\left[
0,\Omega \right] }X\left( d\omega \right) $ with $X\left( d\omega \right)
=X_{11}\otimes \Lambda \left( d\omega \right) $ $+X_{10}\otimes B_{\phi
}^{+}\left( d\omega \right) $ $+X_{01}\otimes B_{\psi }^{-}\left( d\omega
\right) $ $+X_{00}\otimes d\omega $. We can readily obtain the integral
relation 
\begin{eqnarray}
&&\left\langle u\otimes \varepsilon \left( f\right) |\,\left[ X_{\Omega
}Y_{\Omega }-X_{0}Y_{0}-\int_{0}^{\Omega }X_{\omega }Y\left( d\omega \right)
-\int_{0}^{\Omega }X\left( d\omega \right) Y_{\omega }\right] \,v\otimes
\varepsilon \left( g\right) \right\rangle  \notag \\
&=&\int_{[0,\Omega ]}\left\langle u\otimes \varepsilon \left( f\right) |\, 
\left[ X_{11}Y_{11}\left( f_{\omega }|g_{\omega }\right) _{\omega
}+X_{11}Y_{10}\left( f_{\omega }|\phi _{\omega }\right) _{\omega }\right]
\right.  \notag \\
&&\left. \left. +X_{01}Y_{11}\left( \psi _{\omega }|g_{\omega }\right)
_{\omega }+X_{01}Y_{10}\right] \,v\otimes \varepsilon \left( g\right)
\right\rangle \,d\omega  \TCItag{3.12}
\end{eqnarray}
Here we encounter a familiar problem from quantum mechanics: the product of
Wick ordered expressions is not immediately Wick ordered. The non-Leibniz
term in (3.12) is the result of putting to Wick order, in quantum stochastic
calculus it would be called the It\^{o} correction, and for bounded
coefficients we have the It\^{o} product formula 
\begin{equation}
\left( XY\right) \left( d\omega \right) =X\left( d\omega \right) Y\left(
\omega \right) +X\left( \omega \right) Y\left( d\omega \right) +X\left(
d\omega \right) Y\left( d\omega \right)  \tag{3.13}
\end{equation}
which is evaluated by the rule that the fundamental differentials $\Lambda
\left( d\omega \right) ,B_{\phi }^{\pm }\left( d\omega \right) $ and $%
d\omega $ commute with spectrally adapted processes and by the quantum
spectral It\^{o} table: 
\begin{equation}
\begin{tabular}{l|llll}
& $\Lambda \left( d\omega \right) $ & $B_{\phi }^{+}\left( d\omega \right) $
& $B_{\phi }^{-}\left( d\omega \right) $ & $d\omega $ \\ \hline
$\Lambda \left( d\omega \right) $ & $\Lambda \left( d\omega \right) $ & $%
B_{\phi }^{+}\left( d\omega \right) $ & 0 & 0 \\ 
$B_{\psi }^{+}\left( d\omega \right) $ & 0 & 0 & 0 & 0 \\ 
$B_{\psi }^{-}\left( d\omega \right) $ & $B_{\psi }^{-}\left( d\omega
\right) $ & $\left( \psi _{\omega }|\phi _{\omega }\right) _{\omega
}\,d\omega $ & 0 & 0 \\ 
$d\omega $ & 0 & 0 & 0 & 0
\end{tabular}
\tag{3.14}
\end{equation}

\section{Super-symmetric Spectral Transformations}

\noindent \textbf{Definition (4.1):} \emph{We define the spectral parity
processes to be} 
\begin{equation}
J_{\Omega }:=\Gamma _{+}\left( -\Pi _{\left[ 0,\Omega \right] }+\Pi
_{(\Omega ,\infty )}\right) .  \tag{4.1}
\end{equation}
With respect to the decomposition (3.4) we have $J_{\Omega }\equiv \left(
-1\right) ^{\Lambda \left( \Omega \right) }\otimes 1_{(\Omega }$.

\bigskip

\noindent \textbf{Lemma (4.2):} \emph{The process }$\left( J_{\Omega
}\right) _{\Omega \geq 0}$\emph{\ is a unitary, self-adjoint,
frequency-adapted process satisfying the properties} 
\begin{equation*}
\begin{array}{cl}
1. & \left[ J_{\omega },J_{\omega ^{\prime }}\right] =0; \\ 
2. & J_{\omega }\Phi _{+}=\Phi _{+}; \\ 
3. & dJ_{\omega }=-2J_{\omega }\otimes \Lambda \left( d\omega \right)
,J_{0}=1.
\end{array}
\end{equation*}

\begin{proof}
Property 1 follows is immediate from the observation that $J_{\omega
}J_{\omega ^{\prime }}=$ $\Gamma _{+}\left( \Pi _{\lbrack 0,a)}-\Pi
_{\lbrack a,b]}+\Pi _{(b,\infty )}\right) $ where $a=\omega \wedge \omega
^{\prime }$ and $b=\omega \vee \omega ^{\prime }$. Property 2 is evident
from the fact that $\Phi _{+}=\varepsilon \left( 0\right) $.

Next from the rule $\Lambda \left( d\omega \right) \Lambda \left( d\omega
\right) =\Lambda \left( d\omega \right) $, we have 
\begin{equation*}
df\left( \Lambda \left( \omega \right) \right) =\left[ f\left( \Lambda
\left( \Omega \right) +1\right) -f\left( \Lambda \left( \omega \right)
\right) \right] \otimes \Lambda \left( d\omega \right)
\end{equation*}
for analytic functions $f$. Setting $f\left( \lambda \right) =\exp \left\{
i\pi \lambda \right\} $ gives property 3.
\end{proof}

\bigskip

\noindent \textbf{Definition (4.3):} \emph{Let }$\phi \in \frak{h}$\emph{,
we define the Fermionic creation and annihilation spectral processes to be} 
\begin{equation}
F_{\phi }^{\pm }\left( \Omega \right) :=\int_{\left[ 0,\Omega \right]
}J_{\omega }\otimes B_{\phi }^{\pm }\left( d\omega \right) .  \tag{4.2}
\end{equation}

\bigskip

\noindent \textbf{Lemma (4.4):} \emph{For each }$\Omega \geq 0$ \emph{the
Fermionic processes }$F_{\phi }^{\pm }\left( \Omega \right) $\emph{\
anti-commute with the parity operator }$J_{\Omega }$\emph{.}

\begin{proof}
We first note that the exponential vectors are a stable domain for the
parity operator and the Bosonic, and hence Fermionic, processes. In
particular we have 
\begin{gather*}
\left\langle \varepsilon \left( f\right) |\,\left\{ J_{\Omega },F_{\phi
}^{-}\left( \Omega \right) \right\} \,\varepsilon \left( g\right)
\right\rangle =\left\langle \varepsilon \left( f\right) |\,J_{\Omega
}F_{\phi }^{-}\left( \Omega \right) \,\varepsilon \left( g\right)
\right\rangle +\left\langle \varepsilon \left( f\right) |\,F_{\phi
}^{-}\left( \Omega \right) J_{\Omega }\,\varepsilon \left( g\right)
\right\rangle  \\
=\int_{\left[ 0,\Omega \right] }d\omega \,\left( \phi _{\omega }|g_{\omega
}\right) _{\omega }\,\left\langle \varepsilon \left( -\Pi _{\left[ 0,\Omega %
\right] }f+\Pi _{\left( \Omega ,\infty \right) }f\right) |\,\varepsilon
\left( -\Pi _{\left[ 0,\omega \right] }g+\Pi _{\left( \omega ,\infty \right)
}g\right) \right\rangle  \\
-\int_{\left[ 0,\Omega \right] }d\omega \,\left( \phi _{\omega }|g_{\omega
}\right) _{\omega }\left\langle \varepsilon \left( f\right) |\,\varepsilon
\left( -\Pi _{\left[ 0,\Omega \right] }g+\Pi _{\left( \Omega ,\infty \right)
}g\right) \right\rangle  \\
=\int_{\left[ 0,\Omega \right] }d\omega \,\left( \phi _{\omega }|g_{\omega
}\right) _{\omega }\,\left\langle \varepsilon \left( f\right) |\,\left(
J_{\Omega }J_{\omega }-J_{\omega }J_{\Omega }\right) \,\varepsilon \left(
g\right) \right\rangle =0
\end{gather*}
and so we deduce that $\left\{ J_{\Omega },F_{\phi }^{-}\left( \Omega
\right) \right\} =0$. The proof of the relation $\left\{ J_{\Omega },F_{\phi
}^{+}\left( \Omega \right) \right\} $ $=0$ is similar.
\end{proof}

\bigskip

\noindent \textbf{Theorem (4.5):} \emph{The Fermionic processes }$F_{\phi
}^{\pm }\left( \Omega \right) $\emph{\ are bounded and satisfy the canonical
anti-commutation relations} 
\begin{eqnarray}
\left\{ F_{\phi }^{-}\left( \Omega \right) ,F_{\psi }^{+}\left( \Omega
\right) \right\} &=&\int_{\left[ 0,\Omega \right] }d\omega \,\left( \phi
_{\omega }|\psi _{\omega }\right) _{\omega };  \TCItag{4.3a} \\
\left\{ F_{\phi }^{-}\left( \Omega \right) ,F_{\psi }^{-}\left( \Omega
\right) \right\} &=&0=\left\{ F_{\phi }^{+}\left( \Omega \right) ,F_{\psi
}^{+}\left( \Omega \right) \right\} .  \TCItag{4.3b}
\end{eqnarray}

\begin{proof}
Using quantum spectral calculus we have 
\begin{eqnarray*}
d\left\{ F_{\phi }^{-}\left( \omega \right) ,F_{\psi }^{+}\left( \omega
\right) \right\} &=&\left\{ J_{\omega },F_{\psi }^{+}\left( \omega \right)
\right\} \otimes B_{\phi }^{-}\left( d\omega \right) +\left\{ F_{\phi
}^{-}\left( \omega \right) ,J_{\omega }\right\} \otimes B_{\psi }^{+}\left(
d\omega \right) \\
&&+J_{\omega }^{2}\otimes B_{\phi }^{-}\left( d\omega \right) B_{\psi
}^{+}\left( d\omega \right) \\
&=&\left( \phi _{\omega }|\psi _{\omega }\right) _{\omega }d\omega
\end{eqnarray*}
which can be integrated to obtain (4.3a).

Next, since $dF_{\phi }^{-}\left( \omega \right) =J_{\omega }\otimes B_{\phi
}^{+}\left( d\omega \right) $, the It\^{o} formula (3.13) gives 
\begin{equation*}
\left( dF_{\phi }^{-}\left( \omega \right) \right) ^{2}=\left( J_{\omega
}F_{\phi }^{-}\left( \omega \right) +F_{\phi }^{-}\left( \omega \right)
J_{\omega }\right) \otimes B_{\phi }^{-}\left( d\omega \right) =0
\end{equation*}
again by virtue of the previous lemma. Since $F_{\phi }^{-}\left( 0\right)
=0 $, the equation can be integrated to show that $F_{\phi }^{-}\left(
\omega \right) ^{2}=0$. Likewise, we can show that $F_{\phi }^{+}\left(
\omega \right) ^{2}=0$. We then obtain (4.3b) through polarization.
\end{proof}

\bigskip

\noindent \textbf{Lemma (4.6):} \emph{For each }$\Omega \geq 0$ \emph{and }$%
\phi _{1},\dots ,\phi _{n}\in \frak{h}$ \emph{we have the identity} 
\begin{equation*}
F_{\phi _{1}}^{+}\left( \Omega \right) \cdots F_{\phi _{n}}^{+}\left( \Omega
\right) \,\Phi _{+}=\sum_{\sigma \in \frak{S}_{n}}\left( -1\right) ^{\sigma }%
\underset{0<\omega _{1}<\cdots \omega _{n}<\Omega }{\int }B_{\phi _{\sigma
\left( 1\right) }}^{+}\left( d\omega _{1}\right) \cdots B_{\phi _{\sigma
\left( n\right) }}^{+}\left( d\omega _{n}\right) \,\Phi _{+}
\end{equation*}
\emph{where }$\left( -1\right) ^{\sigma }$\emph{\ is the parity of the
permutation }$\sigma \in \frak{S}_{n}$\emph{.}

\begin{proof}
The proof will be one by induction. The identity is trivially true for $n=1$%
.The rules of the spectral quantum stochastic calculus yields 
\begin{equation*}
d\left[ F_{\phi _{1}}^{+}\left( \omega \right) \cdots F_{\phi
_{n+1}}^{+}\left( \omega \right) \right] =
\end{equation*}
\begin{equation*}
\sum_{k=1}^{n+1}\left( -1\right) ^{n-k}\left[ F_{\phi _{1}}^{+}\left( \omega
\right) \cdots \widehat{F_{\phi _{k}}^{+}\left( \omega \right) }\cdots
F_{\phi _{n+1}}^{+}\left( \omega \right) J_{\omega }\right] \otimes B_{\phi
_{k}}^{+}\left( d\omega \right) 
\end{equation*}
therefore 
\begin{gather*}
\left\langle \varepsilon \left( f\right) |\,F_{\phi _{1}}^{+}\left( \Omega
\right) \cdots F_{\phi _{n}}^{+}\left( \Omega \right) \,\Phi
_{+}\right\rangle = \\
\int_{0}^{\Omega }d\omega \,\sum_{k=1}^{n+1}\left( -1\right)
^{n-k}\left\langle \varepsilon \left( f\right) |\,F_{\phi _{1}}^{+}\left(
\omega \right) \cdots \widehat{F_{\phi _{k}}^{+}\left( \omega \right) }%
\cdots F_{\phi _{n+1}}^{+}\left( \omega \right) \,\Phi _{+}\right\rangle
\left( f_{\omega }|\phi _{k}\left( \omega \right) \right) _{\omega };
\end{gather*}
therefore, if the identity holds up to the $n$-th order, then it holds for $%
n+1$ also.
\end{proof}

\bigskip

\noindent \textbf{Corollary (4.7):} \emph{For each }$\Omega \geq 0$ \emph{%
and }$\phi _{1},\dots ,\phi _{n}\in \frak{h}$ \emph{we have the identity} 
\begin{equation*}
B_{\phi _{1}}^{+}\left( \Omega \right) \cdots B_{\phi _{n}}^{+}\left( \Omega
\right) \,\Phi _{+}=\sum_{\sigma \in \frak{S}_{n}}\int_{0<\omega _{1}<\cdots
\omega _{n}<\Omega }F_{\phi _{\sigma \left( 1\right) }}^{+}\left( d\omega
_{1}\right) \cdots F_{\phi _{\sigma \left( n\right) }}^{+}\left( d\omega
_{n}\right) \,\Phi _{+}.
\end{equation*}

\bigskip

\noindent \textbf{Theorem (4.8):} \emph{There exists a unique unitary
mapping }$\Xi :\Gamma _{+}\left( \frak{h}\right) \mapsto \Gamma _{-}\left( 
\frak{h}\right) $\emph{\ with the properties } 
\begin{equation*}
\begin{array}{cl}
1. & \Xi \,\Phi _{+}=\Phi _{-}; \\ 
2. & \Xi \,F_{\phi }^{\pm }\left( \Omega \right) \,\Xi ^{-1}=F^{\pm }\left(
\Pi _{\left[ 0,\Omega \right] }\phi \right) .
\end{array}
\end{equation*}

\begin{proof}
First of all, observe that any $n$-particle vector $\Pi _{+}f_{1}\otimes
\cdots \otimes f_{n}$ can be written as $B^{+}\left( f_{1}\right) \cdots
B^{+}\left( f_{n}\right) \Phi _{+}$ and so any vector in $\Gamma _{+}\left( 
\frak{h}\right) $ can be obtained a sums of products of creators acting on
the Fock vacuum. In other words $\Phi _{+}$ is cyclic in $\Gamma _{+}\left( 
\frak{h}\right) $ for the $B^{+}\left( .\right) $ fields. Likewise, $\Phi
_{+}$ is cyclic for the Fermionic creator fields $F_{\phi }^{+}\left( \Omega
\right) $ too.

We next of all note that the Fermionic annihilator fields $F_{\phi }^{\pm
}\left( \Omega \right) $ annihilate $\Phi _{+}$. We consider the mapping $%
\Xi :\Gamma _{+}\left( \frak{h}\right) \mapsto \Gamma _{-}\left( \frak{h}%
\right) $ defined by linear extension from $\Xi \left( \Phi _{+}\right)
=\Phi _{-}$ and $\Xi \left( F_{\phi _{1}}^{+}\left( \Omega \right) \dots
F_{\phi _{n}}^{+}\left( \Omega \right) \Phi \right) $ $=\Pi _{-}\left( \Pi _{%
\left[ 0,\Omega \right] }\phi _{1}\right) \otimes \dots \otimes \left( \Pi _{%
\left[ 0,\Omega \right] }\phi _{n}\right) $.

It is readily seen that $\Xi $ is a densely defined isometry and so extends
to a unitary.
\end{proof}

\bigskip

\noindent \textbf{Theorem (4.9):} \emph{The unitary mapping }$\Xi :\Gamma
_{+}\left( \frak{h}\right) \mapsto \Gamma _{-}\left( \frak{h}\right) $\emph{%
\ has the following covariance for the differential second quantizations } 
\begin{equation}
\Xi \,\gamma _{+}\left( \Pi _{\left[ 0,\Omega \right] }\right) \,\Xi
^{-1}=\gamma _{-}\left( \Pi _{\left[ 0,\Omega \right] }\right) .  \tag{4.4}
\end{equation}

\begin{proof}
Actually, (4.4) is the differential version of the relation $\Xi \,\Gamma
_{+}\left( U_{t}\right) \,\Xi ^{-1}$ $=\gamma _{-}\left( U_{t}\right) $
where $U_{t}$ is the unitary $\exp \left\{ it\Pi _{\left[ 0,\Omega \right]
}\right\} =e^{it}\Pi _{\left[ 0,\Omega \right] }+\Pi _{(\Omega ,\infty )}$.
This will follow from the fact that $\Gamma _{+}\left( U_{t}\right) \,F^{\pm
}\left( \phi ,\Omega \right) \,\Gamma _{+}\left( U_{t}^{\dagger }\right)
=F^{\pm }\left( U_{t}\phi ,\Omega \right) $ and this is established on the
domain of exponential vectors: 
\begin{eqnarray*}
&&\left\langle \varepsilon \left( f\right) |\,\Gamma _{+}\left( U_{t}\right)
\,F_{\phi }^{\pm }\left( \Omega \right) \,\Gamma _{+}\left( U_{t}^{\dagger
}\right) \,\varepsilon \left( g\right) \right\rangle =\left\langle
\varepsilon \left( U_{t}^{\dagger }f\right) |\,F_{\phi }^{\pm }\left( \Omega
\right) \,\varepsilon \left( U_{t}^{\dagger }g\right) \right\rangle \\
&=&\int_{0}^{\Omega }d\omega \,\left\langle \varepsilon \left(
U_{t}^{\dagger }f\right) |\,J_{\omega }\otimes B_{\phi }^{\pm }\left(
d\omega \right) \,\varepsilon \left( U_{t}^{\dagger }g\right) \right\rangle
\, \\
&=&\left\langle \varepsilon \left( f\right) |\,F_{U_{t}\phi }^{\pm }\left(
\Omega \right) \,\varepsilon \left( g\right) \right\rangle
\end{eqnarray*}
and we stress the importance of the fact that $U_{t}$ is diagonal in our
spectral decomposition.
\end{proof}

\end{document}